\documentstyle[amssymb,aps,twocolumn,epsfig]{revtex}

\newtheorem{theorem}{Theorem}
\newtheorem{acknowledgement}[theorem]{Acknowledgement}

\begin{document}

\twocolumn[\hsize\textwidth\columnwidth\hsize\csname 
@twocolumnfalse\endcsname

\title{Staggered liquid phases of the 1D Kondo-Heisenberg lattice model}
\author{Oron Zachar}
\address{ICTP, 11 strada Costiera, Trieste 34100, Italia}
\date{\today}
\maketitle

\widetext
\begin{abstract}

We introduce a new family of one dimensional liquids, which we label as
''Staggered-Liquids'', in the phase diagram of the 1D Kondo-Heisenberg model. It
encompasses three distinct spin gapped liquids and a new Luttinger liquid
phase. A Staggered Liquid is characterized by gapless modes with ''large
fermi-sea'' signature in the charge density wave (CDW) mode, and in that the
superconducting order involves the near condensation of charge-2e Cooper
pairs with finite center of mass momentum. In particular, the conventional
gapless $2k_{F}$ CDW and $k=0$ pairing modes are absent. In the process, we
analytically derive the phase transition from an intermediate coupling spin
gap phase to the strong coupling gapless LL phase of the Kondo-Heisenberg
lattice model.

\smallskip
\end{abstract}

 ]

\narrowtext

\section{Introduction}

Multicomponent one dimensional electronic systems, of which the one
dimensional Kondo-Heisenberg (K-H) model is a particular example, exhibit
new phases that were unanticipated in the earlier studies of the one
dimensional electron gas (1DEG). The one dimensional Kondo-Heisenberg model
consists of a 1DEG interacting with a Heisenberg spin-$\frac{1}{2}$ chain
via spin exchange interaction. In the present paper we characterize the
stable fixed points of this model.

In particular limits of parameters, we obtain well controlled analytical
solutions which enable us to enumerate and characterize the quantum numbers
of {\em all} gapless modes. Gapless modes are properties of the fixed point.
Thus, our analysis lists the {\em minimal set of stable fixed points in the
global phase diagram} of the Kondo-Heisenberg model. Surprisingly, we find
that there is a common feature to all the fixed points in that CDW and
pairing gapless modes are obtained with ''unusual'' wave numbers, and hence
give the label ''staggered liquids'' to the family of fixed points. In
particular, whereas previously catalogued liquid phases of the 1DEG have a
gapless charge $2e$ pairing mode at $k=0$, in a Staggered Liquid this mode
appears at non-zero wave-vector. Put differently, in a Staggered Liquid
phase the dominant superconducting order involves the near condensation of
Cooper pairs with finite center of mass momentum. Similarly, there is no
gapless CDW mode at wavenumber $2k_{F}$. Our results are summarized in
tables-1,2 of section-II.

In previous publications\cite{zachar-KLL,ZacharTsvelik-KondoHeisenberg}, we
have already characterized two distinct spin gap phases (at weak coupling 
\cite{ZacharTsvelik-KondoHeisenberg} and at a ''Toulouse point'' value of
parameters). In the present paper we add a new gapless Luttinger liquid (LL)
phase (labeled ''Staggered LL) which is obtained by going away from the
Toulouse point towards stronger coupling. Interestingly, though its
mathematical form is similar to commensurate-incommensurate transition, we
find the phase transition is {\em first order}. To our knowledge, it is the
first analytic derivation of the phase transition from an intermediate
coupling spin gap phase to the strong coupling gapless LL phase in the Kondo
lattice model. Moreover, a third distinct spin gap phase is obtained by
introducing weak attractive interactions to the staggered LL (and hence we
name it ''staggered BCS'').

We emphasize that in this paper we limit ourselves to cataloguing the stable
fixed points. We do not discuss the range of their basins of attraction and
the validity of the solutions away from the quantitatively controlled limits
of parameters. In other words, we defer to a future publication the
discussion of how exactly to construct the phase diagram as a function of
Kondo interaction strength for a given discrete Kondo-Heisenberg lattice
model at given incommensurate filling and with spin rotation invariance.

The paper is organized as follows: In section-II, we define the model and
the order parameters. Our results are summarized in tables-1,2. (The
derivation of these results is presented in the ensuing sections). In
section-III, we review the weak coupling limit ($J_{K}\ll J_{H}$) spin gap
fixed point solution\cite{ZacharTsvelik-KondoHeisenberg}. In section-IV, we
review the Toulouse limit ($J_{H}\ll J_{K}\sim E_{F}$) spin gap fixed point
solution, with some extended discussion of the Unitary transformation. In
section-V, we derive the phase transition to a gapless LL away from the
Toulouse point towards stronger coupling ($J_{H}\ll E_{F}\ll J_{K}$). In
section-VI we make some additional concluding remarks. In order to
facilitate the reading of the paper, a discussion and bosonization
representation of order parameters is given in an Appendix.

\section{Kondo-Heisenberg model and its zero temperature fixed points}

\subsection{The model}

The Kondo-Heisenberg model (\ref{H_KondoHeisenberg}) consists of two {\em %
inequivalent} interacting chains; one is a one-dimensional electron gas
(described by the Hamiltonian $H^{1DEG}$\cite{1D-ref}), and the other an
antiferromagnetic Heisenberg chain of localized spins 1/2, $\left\{ \vec{\tau%
}_{j}\right\} $. The chains interact via a spin exchange interaction with an
antiferromagnetic coupling constant $J_{K}>0$.

\begin{eqnarray}
H &=&H^{1DEG}+H^{Heis}+H_{K}{\bf \ }  \label{H_KondoHeisenberg} \\
H^{Heis} &=&J_{H}\sum_{j}{\bf \vec{\tau}}_{j}\cdot {\bf \vec{\tau}}_{j+1}~~
\\
H_{K} &=&2J_{K}\sum_{j}{\bf \vec{\tau}}_{j}\cdot \vec{s}\left( x_{j}\right)
\end{eqnarray}
where ${\bf \vec{s}}\left( x_{j}\right) =\psi _{\alpha }^{\dagger }(x_{j})%
\frac{{\bf \sigma }_{\alpha \beta }}{2}\psi _{\beta }(x_{j})$ is the
electron gas spin density operator at position $x_{j}$ of the local spin $%
{\bf \vec{\tau}}_{j}$ of the Heisenberg chain. We focus on the low energy
and long-distance behavior of the electron's correlation functions by taking
the continuum limit of the electron gas and linearizing the 1DEG dispersion
relation about the fermi points, $\pm k_{F}$, with corresponding right and
left going electron fields, $R_{\sigma }$ and $L_{\sigma }$; 
\[
\psi _{\sigma }\left( x\right) =R_{\sigma }(x)e^{+ik_{F}x}+L_{\sigma
}(x)e^{-ik_{F}x}, 
\]
Where $\sigma =\uparrow ,\downarrow $.

\begin{figure}
\begin{center}
\leavevmode
\epsfxsize=3in \epsfbox{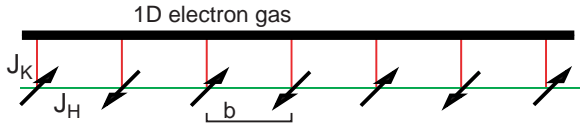}
\end{center}
\caption{Kondo-Heisenberg model}
\label{K-H model-fig}
\end{figure}

The effective Fermi wave numbers (in the sense of the generalized
Luttinger's theorem \cite{Generalized-Luttinger}) for the 1DEG and the spin
chain are $2k_{F}$ and $2k_{F}^{Heis}=\pi /b$ respectively (where $%
b=x_{j+1}-x_{j}$ is the distance between the local spins of the Heisenberg
chain). It is assumed that the two systems are mutually incommensurate, and
that $2k_{F}$ is incommensurate with any underlying ionic lattice. The
continuum limit is taken for the 1DEG while the Heisenberg chain is
initially left discrete (and remain so in some of the limit solutions
derivations). Therefore, {\em the totality of our analysis is rigorously
valid in the limit} $2k_{F}\gg \pi /b$ (i.e., the number of electrons is
much larger than the number of local spin-$\frac{1}{2}$ moments per unit
length).

The 1DEG spin currents are decomposed into forward and back-scattering
parts; 
\begin{eqnarray*}
{\bf s}\left( x\right) &=&\psi _{\alpha }^{\dagger }(x_{j})\frac{{\bf \vec{%
\sigma}}_{\alpha \beta }}{2}\psi _{\beta }(x_{j}) \\
&=&\vec{J}_{s}\left( x\right) +\vec{n}_{s}\left( x\right)
\end{eqnarray*}
where $\vec{J}_{s}\left( x\right) {\bf =}\vec{J}_{sR}\left( x\right) +\vec{J}%
_{sL}\left( x\right) $, $\vec{J}_{sR}=\frac{1}{2}R_{\sigma }^{+}\vec{\sigma}%
_{\sigma \sigma ^{\prime }}R_{\sigma ^{\prime }}$ $;$ $\vec{J}_{sL}=\frac{1}{%
2}L_{\sigma }^{+}\vec{\sigma}_{\sigma \sigma ^{\prime }}L_{\sigma ^{\prime
}} $ are the ferromagnetic ($q=0$) spin currents of right- and left-moving
electrons respectively, and 
\[
\vec{n}_{s}\left( x\right) =e^{-i2k_{F}x_{j}}\vec{n}_{R}\left( x\right)
+e^{+i2k_{F}x_{j}}\vec{n}_{L}\left( x\right) 
\]
where $\vec{n}_{R}=R_{\sigma }^{+}\frac{\vec{\sigma}_{\sigma ,\sigma
^{\prime }}}{2}L_{\sigma ^{\prime }}$ $;$ $\vec{n}_{L}=L_{\sigma }^{+}\frac{%
\vec{\sigma}_{\sigma ,\sigma ^{\prime }}}{2}R_{\sigma ^{\prime }}$ are the
staggered magnetization ($q=2k_{F}$) components of the 1DEG.

Due to the incommensurate electron filling, back-scattering interaction
terms are irrelevant in the renormalization group (RG) sense, and for our
purposes may be dropped from the Hamiltonian which describe the fixed
points. As a result, the spin and charge sectors decouple, 
\[
H=\int dx\left[ {\cal H}_{c}+{\cal H}_{spin}\right] . 
\]
The charge sector is described by a Gaussian model\cite{1D-ref} 
\[
{\cal H}_{c}=\frac{1}{2}\left[ K_{c}\Pi _{c}^{2}\left( x\right) +\frac{1}{%
K_{c}}v_{c}\left( \partial _{x}\Phi _{c}\right) ^{2}\right] . 
\]
Since in this paper we are not interested in the effects of anomalous 1D
exponents ($K_{c}\neq 1$), we will set $K_{c}=1$, unless otherwise
explicitly stated. The subsequent analysis and manipulations deal only with
the spin sector fields. The Kondo exchange interaction reduces to 
\begin{equation}
H_{K}=J_{K}\sum_{j}{\bf \vec{\tau}}_{j}\cdot \vec{J}\left( x_{j}\right) .
\end{equation}

\subsection{Order parameters and staggered correlations}

Study of the different stable phases of the Kondo-Heisenberg array begins
with an analysis of the gapless excitations of the decoupled fixed point.
From there, as usual, we sort the phases by determining which of these
excitations become gapped, and which remain gapless in the presence of the
(Kondo) couplings between the 1DEG and the Heisenberg chain. Since our
ultimate goal is to study the coupled system, we need also to consider the
character of gapless excitations constructed of composites operators from
the two subsystems. An extensive exposition of the order parameters is given
in the appendix. Below, we note only the modes which are relevant for a spin
gap system.

In the spin gap phases, only spin-$0$ modes may be gapless. Thus, we focus
our investigation on singlet pairing modes (charge-$2e$, spin-$0$) and CDW
modes (charge-$0$, spin-$0$). The corresponding usual 1DEG order parameters
are:

\begin{eqnarray}
{\cal O}_{SP} &=&\frac{1}{\sqrt{2}}\left( R_{\uparrow }^{\dagger
}L_{\downarrow }^{\dagger }+L_{\uparrow }^{\dagger }R_{\downarrow }^{\dagger
}\right) \\
{\cal O}_{CDW} &=&\frac{1}{2}\left[ \left( R_{\uparrow }^{\dagger
}L_{\uparrow }+R_{\downarrow }^{\dagger }L_{\downarrow }\right) +{\rm h.c.}%
\right] .
\end{eqnarray}
Modes of {\em composite} nature are: A composite odd-parity singlet pairing 
\begin{equation}
{\cal O}_{c-SP}=-i\left[ R_{\alpha }^{\dagger }\left( {\bf \vec{\sigma}}%
\sigma _{2}\right) _{\alpha \beta }L_{\beta }^{\dagger }\right] \cdot {\bf 
\vec{\tau}},
\end{equation}
and a composite particle-hole mode, ${\cal O}_{c-CDW}$, which will play a
central role in the ensuing discussion 
\begin{equation}
{\cal O}_{c-CDW}=\vec{n}_{1DEG}\cdot {\bf \vec{\tau}.}
\end{equation}

Upon evaluating the corresponding correlation functions $\chi
_{i}(x,x^{\prime })=\left\langle {\cal O}_{i}(x){\cal O}_{i}(x^{\prime
})\right\rangle $, we find gapless modes with power law correlations of the
form. 
\begin{equation}
\chi _{i}\left( x_{j}-x_{j^{^{\prime }}}\right) =\left( -1\right) ^{\left(
j-j^{\prime }\right) }\chi _{0}\left( x_{j}-x_{j^{^{\prime }}}\right) ,
\label{stagger}
\end{equation}
where $\chi _{0}\sim x^{-\alpha _{i}}$ . The staggering factor $\left(
-1\right) ^{j}$ in the correlation functions (\ref{stagger}) is effectively
modulating the usual power law correlations by the reciprocal lattice
vector, $\frac{\pi }{b}$, of the spin chain $\left\{ {\bf \tau }_{j}\right\} 
$. As a result, the gapless modes are found in unusual finite momentum
values: The singlet pairing are with momentum $\frac{\pi }{b}$ (and there is
no $k=0$ singlet pairing with charge $2e$), and the gapless $CDW$ modes are
at momentum 
\begin{equation}
2k_{F}^{\ast }=2k_{F}+\frac{\pi }{b}  \label{2k_F*}
\end{equation}
(and not at $2k_{F}$ as CDW in free 1DEG). These are {\em the defining
characteristics of a ''staggered liquid''}.

Insight into the gapless modes properties is gained by considering the so
called $\eta $-pairing modes at momentum $\pm 2k_{F}$, 
\begin{eqnarray}
\eta _{R} &=&R_{\uparrow }^{\dagger }R_{\downarrow }^{\dagger }
\label{eta-pair} \\
\eta _{L} &=&L_{\uparrow }^{\dagger }L_{\downarrow }^{\dagger }  \nonumber
\end{eqnarray}
corresponding to right and left going singlet pairs. In a bosonization
representation (Appendix-A) it is easy to see that the $\eta $-pairing
operators depend only on the 1DEG charge sector fields. The charge sector is
unaffected by the relevant part of the Kondo and Heisenberg interactions in
all the zero temperature fixed point. Therefore, the gapless $\eta $-pairing
modes always exist and carry momentum $2k_{F}$ as in free 1DEG. It is
instructive to define operators 
\begin{eqnarray}
\eta ^{even} &\equiv &\frac{1}{\sqrt{2}}\left( \eta _{R}+\eta _{L}\right) \\
\eta ^{odd} &\equiv &\frac{1}{\sqrt{2}}\left( \eta _{R}-\eta _{L}\right) . 
\nonumber
\end{eqnarray}
(thought in themselves they do not carry a well defined momentum quantum
number). We found that an interdependence of the gapless modes is
established by the following operator identities; 
\begin{eqnarray}
{\cal O}_{SP} &=&\left[ {\cal O}_{CDW},\eta ^{even}\right]
\label{[CDW,eta-even]} \\
0 &=&\left[ {\cal O}_{CDW},\eta ^{odd}\right]  \label{[CDW,eta-odd]} \\
0 &=&\left[ {\cal O}_{c-CDW},\eta ^{even}\right]  \label{[c-CDW,eta-even]} \\
{\cal O}_{c-SP} &=&\left[ {\cal O}_{c-CDW},\eta ^{odd}\right] .
\label{[c-CDW,eta-odd]}
\end{eqnarray}
Hence, the {\em gapless wavenumbers of CDW and pairing operators are always
connected by momentum} $2k_{F}$.

\subsection{Main results: Staggered liquids fixed points}

For the purpose of characterizing fixed points, the issue of counting
gapless modes requires clarification. Since the $\eta $-pairing modes are
gapless in all cases were the charge sector is gapless (i.e., at all the
fixed points of the Kondo-Heisenberg lattice model at incommensurate
filling), the interdependence of modes (given in equations \ref
{[CDW,eta-even]},\ref{[c-CDW,eta-odd]}) implies that formally only the CDW
modes need to be counted, while the pairing modes ${\cal O}_{SP}$ and ${\cal %
O}_{c-SP}$ are redundantly derived from combinations of CDW and $\eta $%
-pairing operators. In spite of that, since common discussions in the
literature are done in terms of the usual pairing order parameters ${\cal O}%
_{SP}$ and ${\cal O}_{c-SP}$, we will list them in our tables below.

We have found four distinct fixed points associated with different parameter
values:

\begin{enumerate}
\item  $J_{K}\ll J_{H}$ $\ll E_{F}$ ; A spin gap phase at weak-intermediate
coupling\cite{ZacharTsvelik-KondoHeisenberg}.

\item  $J_{K}\gtrsim E_{F}\gg J_{H}$ ; A spin gap ''Toulouse point'' phase 
\cite{zachar-KLL} at intermediate coupling.

\item  $J_{K}\gg E_{F}\gg J_{H}$ ; A gapless {\sl staggered} Luttinger
liquid phase at strong coupling.

\item  $J_{K}\gg E_{F}\gg J_{H}$ ; A spin gap {\sl staggered} BCS phase at
strong coupling, with additional weak attractive charge interactions.
\end{enumerate}

We comment that all the fixed point Hamiltonians (associated with the above
noted phases) which we derived are in fact spin rotation invariant, even
though the bare interaction parameters were in some cases breaking spin
rotation invariance. It is an example of the possibility that the ultimate
zero temperature fixed point can posses higher symmetry than the original
microscopic model. Yet, we emphasize again that such issues do not affect
the validity of our analysis for cataloging the fixed points of the most
general microscopic Kondo-Heisenberg model (with or without spin rotation
invariance).

In the tables below, we characterize the above noted fixed points in terms
of momentum quantum number of their gapless CDW and pairing modes (''X''
signifies that the particular mode is gapped).

Spin gap phases fixed points: 
\[
\stackrel{\text{Table-1: Spin gap phases}}{
\begin{tabular}{|ccccc|}
\hline
& ${\cal O}_{CDW}$ & $\ \ \ {\cal O}_{SP}$ \  & ${\cal O}_{c-CDW}$ & $\ 
{\cal O}_{c-SP}$ \\ \hline
$
\begin{array}{c}
\text{weak} \\ 
\text{coupling}
\end{array}
$ & X & X & $2k_{F}^{\ast }$ & $\frac{\pi }{b}$ \\ \hline
$
\begin{array}{c}
\text{Toulouse} \\ 
\text{point}
\end{array}
$ & $2k_{F}^{\ast }$ & $\frac{\pi }{b}$ & $2k_{F}^{\ast }$ & $\frac{\pi }{b}$
\\ \hline
$
\begin{array}{c}
\text{Staggered} \\ 
\text{BCS}
\end{array}
$ & $2k_{F}^{\ast }$ & $\frac{\pi }{b}$ & X & X \\ \hline
\end{tabular}
} 
\]
Gapless Luttinger liquid fixed points: 
\[
\stackrel{\text{Table-2: Gapless Luttinger liquid phase}}{
\begin{tabular}{|ccccc|}
\hline
& ${\cal O}_{CDW}$ & $\ \ \ {\cal O}_{SP}$ \  & ${\cal O}_{c-CDW}$ & $\ \ 
{\cal O}_{c-SP}$ \\ \hline
$
\begin{array}{c}
\text{Staggered} \\ 
\text{LL}
\end{array}
$ & $2k_{F}^{\ast }$ & $\frac{\pi }{b}$ & $2k_{F}^{\ast }$ & $\frac{\pi }{b}$
\\ \hline
\end{tabular}
} 
\]
Obviously, the gapless LL is characterized by having also gapless spin
density wave (SDW) modes and triplet pairing modes.

Whereas previously catalogued liquid phases of the 1DEG have a gapless
charge $2e$ pairing mode at $k=0$, in all of the above noted phases this
mode appears at non-zero wave-vector. Similarly, there is no gapless CDW
mode at wavenumber $2k_{F}$.

\section{Weak coupling limit ($J_{K}\ll J_{H}$) spin gap fixed point}

In the {\em weak inter-chain coupling limit} 
\[
J_{K}\ll J_{H},E_{F}. 
\]
It is allowed to make {\em further approximation} by taking the continuum
limit also for the Heisenberg spin chain (such approximation is not valid in
the opposite limit $J_{K}\gg J_{H}$, which is discussed in section-IV). The
local spin chain field is then also decomposed into the smooth
(ferromagnetic) and staggered (antiferromagnetic) components; 
\begin{equation}
{\bf \vec{\tau}}_{j}=\left[ {\bf J}_{R}^{\tau }\left( x_{j}\right) +{\bf J}%
_{L}^{\tau }\left( x_{j}\right) \right] +\left( -1\right) ^{j}{\bf n}_{\tau
}\left( x_{j}\right) .  \label{Impurity-spin}
\end{equation}
(Note: we will consistently use the subscripts ''$\tau ,s$'' to distinguish
the spin chain fields from the 1DEG fields).

In order to distinguish contributions coming from various interaction terms,
we introduce distinct Kondo coupling coefficients for forward scattering ($%
J_{f}$) and mixed interactions ($J_{m}$); 
\begin{eqnarray}
H_{K} &=&J_{f}\left( {\bf J}_{R}^{\tau }+{\bf J}_{L}^{\tau }\right) \cdot
\left( {\bf J}_{R}^{s}+{\bf J}_{L}^{s}\right)   \label{J_K-components} \\
&&+J_{m}\left( -1\right) ^{j}{\bf n}_{\tau }\cdot \left( {\bf J}_{R}^{s}+%
{\bf J}_{L}^{s}\right)   \nonumber
\end{eqnarray}
The mixed interaction, of the ferromagnetic 1DEG component with the {\it %
staggered impurity component} (i.e., the $J_{m}\left( -1\right) ^{j}{\bf n}%
_{\tau }\cdot \left( {\bf J}_{R}+{\bf J}_{L}\right) $ term) has naive
scaling dimension $\frac{3}{2}$, but the oscillating $\left( -1\right) ^{j}$
factor, which acts as an effective extra derivative factor $\left( \partial
_{x}\right) $, renders this term to be perturbatively irrelevant in the
renormalization group sense with respect to the free Hamiltonian, $H_{0}^{s}$%
. The forward current-current interaction, $J_{f}\left( {\bf J}_{\tau R}+%
{\bf J}_{\tau L}\right) \cdot \left( {\bf J}_{R}+{\bf J}_{L}\right) $, has
scaling dimension $2$ and is marginal relevant and leads to the opening of a
spin gap \cite{ZacharTsvelik-KondoHeisenberg,Affleck-zigzag}. (The{\sl \ }$%
J_{m}$ term will prove to be essential for understanding the Toulouse limit
solution in section-IV). Therefore, at incommensurate filling in the weak
coupling limit, the Kondo-Heisenberg Hamiltonian (\ref{H_KondoHeisenberg})
reduces to 
\begin{eqnarray}
H_{weak} &=&{\cal H}^{c}+{\cal H}_{0}^{s}+J_{f}\int {\rm d}x\left( {\bf J}%
_{R}^{\tau }+{\bf J}_{L}^{\tau }\right) \cdot \left( {\bf J}_{R}^{s}+{\bf J}%
_{L}^{s}\right)   \label{H* - WeakCoupling} \\
{\cal H}^{c} &=&\frac{1}{2}\left[ K\Pi _{c}^{2}\left( x\right) +\frac{1}{K}%
v_{c}\left( \partial _{x}\Phi _{c}\right) ^{2}\right]   \nonumber \\
{\cal H}_{0}^{s} &=&\sum_{\mu =s,\tau }\frac{2\pi v_{\mu }}{3}\left( :{\bf J}%
_{R}^{\mu }{\bf J}_{R}^{\mu }:+:{\bf J}_{L}^{\mu }{\bf J}_{L}^{\mu }:\right) 
\nonumber
\end{eqnarray}
where $v_{\tau },v_{s}$ are the spin wave velocities of the Heisenberg chain
and 1DEG respectively ($v_{\tau }=\pi J_{{\rm H}}/2$). For a detailed
derivation of the gapless modes of model (\ref{H* - WeakCoupling}), we refer
the reader to our paper\cite{ZacharTsvelik-KondoHeisenberg}. The end results
are quoted in the first line of table-1. It is remarkable that {\em only}
composite modes are gapless.

\section{Toulouse limit ($J_{H}\ll J_{K}\sim E_{F}$) spin gap fixed point}

In the limit 
\[
J_{H}\ll J_{K}\sim E_{F} 
\]
the intra-chain interaction, $J_{H}$, is small compared with the interchain
interaction, $J_{K}$, and it is {\em incorrect} to take the continuum limit
for the spin chain prior to accounting for the effect of the interaction $%
J_{K}$. For simplicity, since $J_{H}\ll J_{K}$, we will first take the limit 
$J_{H}=0$. (We shall find that bringing back $J_{H}\ll J_{K}$ is an
irrelevant perturbation, in the renormalization group sense, due to the spin
gap of ''Toulouse fixed point'' phase). Thus, we model the local spins as
initially independent, and {\em leave the Kondo interaction in its discrete
form}. 
\begin{equation}
H=H_{0}^{1DEG}+2J_{K}\sum_{j}{\bf \tau }_{j}\cdot \psi _{\alpha }^{\dagger
}(x_{j})\frac{{\bf \sigma }_{\alpha \beta }}{2}\psi _{\beta }(x_{j}){\bf \ }
\label{H-Kondo}
\end{equation}
In this limit, effective interaction and coherence between the local spins
will come about explicitly mediated by the itinerant 1DEG (i.e., in a kind
of RKKY which is {\em not} introduce by hand to the model as $J_{H}$).

Below, we review and discuss the ''Toulouse point'' derivation and results
of \cite{zachar-KLL}. For the purpose of calculating correlation functions,
we bosonize the 1DEG fermionic fields\cite{1D-ref}; 
\begin{eqnarray*}
L_{\sigma }\left( x\right) &=&\frac{F_{\sigma }}{\sqrt{2\pi a}}e^{-i\sqrt{%
\pi }\left[ \theta _{\sigma }\left( x\right) +\phi _{\sigma }\left( x\right) %
\right] } \\
R_{\sigma }\left( x\right) &=&\frac{F_{\sigma }}{\sqrt{2\pi a}}e^{-i\sqrt{%
\pi }\left[ \theta _{\sigma }\left( x\right) -\phi _{\sigma }\left( x\right) %
\right] }
\end{eqnarray*}
Where $\theta _{\sigma }(x)=\int_{-\infty }^{x}dx^{\prime }\Pi _{\sigma
}(x^{\prime })$, and $\left[ \Pi _{\sigma }(x^{\prime }),\phi _{\sigma }(x)%
\right] =-i{\delta }(x^{\prime }-x)$, $\sigma =\uparrow ,\downarrow $. The
anticommuting Klein factors, $\{F_{\sigma },F_{\sigma ^{\prime }}\}=\delta
_{\sigma ,\sigma ^{\prime }}$, are needed for the proper anticommutation of
fermions with different spin. As commonly done, we re-express the operators
in terms of bosonic spin fields $\phi _{s}(x)=\frac{1}{\sqrt{2}}[\phi
_{\uparrow }-\phi _{\downarrow }]$, and charge fields $\phi _{c}(x)=\frac{1}{%
\sqrt{2}}[\phi _{\uparrow }+\phi _{\downarrow }]$, and correspondingly
defined momenta $\Pi _{s}$ and $\Pi _{c}$.

The crucial new step which we introduced in \cite{zachar-KLL} is to make a
unitary transformation of the fields, 
\begin{eqnarray}
&&U=\exp \left[ {-i\sqrt{2\pi }\sum_{j}\tau _{j}^{z}\theta _{s}(x_{j})}%
\right] ,  \label{U} \\
&&U\sqrt{2\pi }\left( \partial _{x}\phi _{s}\right) U{^{\dag }}=\sqrt{2\pi }%
\left( \partial _{x}\phi _{s}\right) -2\pi \sum_{l}\tau _{l}^{z}\delta
(x_{l}-x)  \label{U-trans1} \\
&&U\tau ^{+}e^{-i\sqrt{2\pi }\theta _{s}}U{^{\dag }}{=}\tau ^{+}.
\label{U-trans2} \\
&&U\cos \left[ \sqrt{2\pi }\phi _{s}\left( j\right) \right] U{^{\dag }}{=}%
\left( -1\right) ^{j}\cos \left[ \sqrt{2\pi }\phi _{s}\left( j\right) \right]
\label{U-trans cos()}
\end{eqnarray}
In words, going across an impurity, the spin-phase $\sqrt{2\pi }\phi _{s}$
is shifted by $\pm \pi $. i.e., transformed fields with opposite spins
acquires a phase shift of $\delta _{\sigma }=\pm \frac{\pi }{2}$. It is
reminiscent of the unitarity limit scattering we expect from the low energy
physics of the single impurity Kondo effect\cite{Nozieres}. Therefore, we
interpret the unitary transformation as going to a Kondo strong coupling
basis.

The resulting transformed Kondo lattice Hamiltonian is given in equation (%
\ref{H-trans}),  
\begin{eqnarray}
U^{\dagger }HU &=&\tilde{H}_{0}+\Delta J_{z}\sqrt{\frac{2}{\pi }}%
\sum_{j}\tau _{j}^{z}\partial _{x}\phi _{s}\left( x_{j}\right) 
\label{H-trans} \\
&&+\frac{J_{\perp }}{\pi a}\sum_{j}\tau _{j}^{x}(-1)^{j}\cos [\sqrt{2\pi }%
\phi _{s}(j)]  \nonumber \\
\tilde{H}_{0} &=&H_{0}^{s}+H_{0}^{c}-\left( J_{z}+\Delta J_{z}\right) \frac{1%
}{b}\sum_{j}\left( \tau _{j}^{z}\right) ^{2}  \label{H0-trans}
\end{eqnarray}
where, $H_{0}^{s}=\frac{v_{s}}{2}\int dx\left[ \Pi _{s}^{2}+\left( \partial
_{x}\phi _{s}\right) ^{2}\right] $, and 
\begin{equation}
\Delta J_{z}=J_{z}-\pi v_{F}.
\end{equation}
In (\ref{H-trans}) we have introduced independent interaction coefficient $%
J_{z}$ and $J_{\perp }$ for the Ising and spin-flip parts of the Kondo
exchange interaction $J_{K}$. Hence, formally we are here examining a
generalization of the Kondo-Heisenberg model (\ref{H_KondoHeisenberg}) to
non spin-rotation-invariant interactions.

The {\em transformed fields constitute the low energy spectrum of }$\tilde{H}%
_{0}${\em \ into which part of the interaction energy has been incorporated}%
. The transformed fields are taking advantage of the Ising part of the
magnetic Kondo interaction at the cost of kinetic energy (due to twisting of
the spin field $\phi _{s}\left( x\right) $). These are originally high
energy states of the bare free 1DEG Hamiltonian, $H^{1DEG}$. For the
transformed fields to become the new low energy states due to interactions,
it is clear that the Kondo interaction strength needs to be on the order of
the 1DEG bandwidth. To this effect, note the shift of the ground-state
energy per impurity (irrespective of the existence of a spin gap) in
equation (\ref{H0-trans}); 
\begin{eqnarray}
\Delta E_{j} &=&-\left( J_{z}+\Delta J_{z}\right) \frac{1}{b}\sum_{j}\left(
\tau _{j}^{z}\right) ^{2}  \label{U_energy-shift} \\
&=&-\left( 2J_{z}-\pi v_{F}\right) \frac{1}{4b}.  \nonumber
\end{eqnarray}
It represents the absorption of a part of the Kondo interaction energy, $-%
\frac{2J_{z}}{4b}$ (equal to the gain from forming an {\it Ising singlet})
into the transformed free field Hamiltonian (\ref{H0-trans}), at the cost of
kinetic energy $+\frac{\pi v_{F}}{4b}$. Hence, for strong enough
interactions the transformed free fields have lower energy than the bare
1DEG free fields, and therefore determine the low frequency correlations of
various order parameters. Thus, the Toulouse point solution is an outcome of
finite ''strong enough'' interactions and cannot be reached by perturbative
methods about the non-interacting basis.

For a special value of the coupling constants, 
\begin{equation}
J_{z}=\pi v_{F}\Longrightarrow \Delta J_{z}=0
\label{Toulouse point condition}
\end{equation}
(the ''Toulouse point''), we are left with an exactly solvable fixed point
Hamiltonian \cite{zachar-KLL}, 
\begin{equation}
\tilde{H}^{\ast }=H_{0}^{c}+H_{0}^{s}+\frac{J_{\perp }^{f}}{\pi a}%
\sum_{j}\tau _{j}^{x}(-1)^{j}\cos [\sqrt{2\pi }\phi _{s}(j)].
\label{H* - Toulouse}
\end{equation}
The spin part of the fixed point Hamiltonian has a discrete sine-Gordon
form, and therefore, a spin gap. The {\em transformed} spin fields which
develope an expectation value are{\sl \ }$\left\langle \tau
_{j}^{x}(-1)^{j}\right\rangle \neq 0$ and $\left\langle \cos [\sqrt{2\pi }%
\phi _{s}(j)]\right\rangle \neq 0$. In calculating correlation functions, it
is important to remember the effect of the unitary transformations which
lead to 
\[
\left\langle \cos [\sqrt{2\pi }\phi _{s}(x)]\cos [\sqrt{2\pi }\phi
_{s}(x^{\prime })]\right\rangle \sim \left( -1\right) ^{j-j^{\prime }} 
\]
($j(x)$ is defined as the $j$-impurity site to the left of position $x$).
The bare impurity correlations $\left\langle \tau _{j}^{x}\tau _{j^{^{\prime
}}}^{x}\right\rangle $ decays exponentially. However, the transformed
impurity spin, $\tilde{\tau}_{j}^{x}=U{^{\dag }}\tau _{j}^{x}U$, exhibit
staggered long-range order at $T=0$, $\left\langle \tilde{\tau}_{j}^{x}%
\tilde{\tau}_{j^{\prime }}^{x}\right\rangle =const\cdot (-1)^{(j-j^{\prime
})}$; This non-local order parameter characterizes the coherent ground
state. That is all the information needed for deducing the correlation
functions of all order parameters, and thus determine the gapless modes as
done in\cite{zachar-KLL} and summarized in the second line of table-1.

We take this opportunity to elaborate on the significance of the fields
transformation. The $\pi $ phase shift of the field $\sqrt{2\pi }\phi _{s}$
across an impurity site (\ref{U-trans1}), give rise to a staggered
coefficient $(-1)^{j}$ in the Hamiltonian (\ref{H* - Toulouse}) since 
\[
\exp \left[ i2\pi \sum_{l=1}^{j}\tau _{l}^{z}\right] =\left( -1\right) ^{j}. 
\]
Note that the factor $\left( -1\right) ^{j}$ is effectively ''counting''
impurities, and is obtained irrespective of the order of the bare $\left\{
\tau _{l}^{z}\right\} $ themselves (imagine an Ising chain of $\left\{ \tau
_{l}^{z}\right\} $, there is a factor $e^{\pm i\pi }=-1$ per impurity).
Indeed, the correlation function $\left\langle \tau _{j}^{z}\tau _{j^{\prime
}}^{z}\right\rangle $ is short range.

It is interesting to trace back the relevant interaction in the Toulouse
fixed point Hamiltonian in terms of the continuum limit of the Heisenberg
spin chain (\ref{J_K-components}). Due to the additional $\left( -1\right)
^{j}$ phase factor, in the transformed basis, the relevant slowly varying
interaction is now $J_{m}{\bf n}_{\tau }\cdot \left( {\bf J}_{R}+{\bf J}%
_{L}\right) $, while the interaction $J_{f}\left( -1\right) ^{j}\left( {\bf J%
}_{\tau R}+{\bf J}_{\tau L}\right) \cdot \left( {\bf J}_{R}+{\bf J}%
_{L}\right) $ is now also rapidly oscillating and irrelevant!! Thus,
Toulouse fixed point physics originates from the interaction $J_{m}$ that\
couples the conduction electrons to the staggered component of the impurity
array, an interaction that is relevant only with respect to the transformed
fixed point Hamiltonian, $\tilde{H}_{0}$, and was irrelevant in the
untransformed basis. This possibility would be missed in the continuum limit
if we had dropped the $J_{m}\left( -1\right) ^{j}{\bf n}_{\tau }\cdot \left( 
{\bf J}_{R}+{\bf J}_{L}\right) $ term at the outset (as is usually done,
e.g., in \cite{Affleck-zigzag}). The perturbative relevance of various
interaction terms is changed after a transformation to the ''proper'' strong
coupling basis of fields about which perturbative RG analysis is performed.
The notation $J_{m}$ is not accidental, and it is exactly the one which is
responsible for the non-trivial fixed point of the two-impurity Kondo problem
\cite{Gan95-Kondo2impurities}.

\section{Stronger coupling ($J_{H}\ll E_{F}\ll J_{K}$) staggered Luttinger
liquid fixed point}

\subsection{Phase transition away from the Toulouse limit}

In a previous paper \cite{zachar-KLL}, we analyzed the
commensurate-incommensurate (C-I) transition in the {\em charge} sector {\em %
at} the Toulouse point, as a function of the filling factor, and found a
phase transition from an insulating phase (with both charge and spin gaps)
to a conducting phase with only a spin gap. Here, we are interested only in
the case of incommensurate filling (for which there is no charge gap). In
this section, we analyze the phase transitions in the {\em spin} sector by
varying the parameter values {\em away from} the Toulouse point, While
maintaining the same incommensurate charge filling factor.

We investigate the {\em phase transitions within the transformed fields
Hamiltonian }(\ref{H-trans}). The local stability of the Toulouse limit
fixed point ($J_{z}^{\ast }=\pi v_{F}$) is ensured by the existence of a
spin-gap. This is all that can be deduced from perturbative renormalization
group calculations. Thus, the phase transition can be established only via
non-perturbative methods. Below, we determine analytically the finite
parameter space region characterized by the Toulouse fixed point solution.
i.e., the zero temperature stability of the spin gap to finite deviations $%
\Delta J_{z}=\left( J_{z}-\pi v_{F}\right) >0$ away from the Toulouse line
towards stronger coupling. We find a new electronic gapless phase beyond a
finite distance from the Toulouse point.

Treating the transformed impurity spins in self-consistent mean-field
approximation, we replace them by their expectation values in the
transformed Hamiltonian; 
\begin{eqnarray}
U^{\dagger }HU\rightarrow H_{0}^{s}+H_{0}^{c} &&+\Delta J_{z}\sqrt{\frac{2}{%
\pi }}\sum_{j}\left\langle \tau _{j}^{z}\right\rangle \partial _{x}\phi
_{s}(x_{j})  \label{H-MF} \\
&&+\frac{J_{\perp }}{\pi a}\sum_{j}\left\langle (-1)^{j}\tau
_{j}^{x}\right\rangle \cos [\sqrt{2\pi }\phi _{s}(j)].  \nonumber
\end{eqnarray}
The {\em spin sector} of the Hamiltonian (\ref{H-MF}) has a form familiar
from the study of Commensurate-Incommensurate (C-I) transitions;

\begin{eqnarray}
\frac{\tilde{H}^{s}}{v_{s}} &=&\frac{1}{2}\int dx\left[ \Pi
_{s}^{2}+(\partial _{x}\phi _{s}-\delta )^{2}\right]  \label{H-CI} \\
&&+h\int dx\cos [\beta \phi _{s}\left( x\right) ]  \nonumber
\end{eqnarray}
Where, $\beta =\sqrt{2\pi }$, 
\begin{eqnarray}
\delta &=&\delta _{0}\sin \left( \gamma \right) =\Delta J_{z}\frac{c}{a}%
\sqrt{\frac{2}{\pi }}\left\langle \tau _{j}^{z}\right\rangle
\label{mean field coefficients} \\
h &=&h_{0}\cos \left( \gamma \right) =J_{\perp }\left| \left\langle \tau
_{j}^{x}\right\rangle \right| \frac{c}{2\pi a^{2}v_{s}}.  \nonumber
\end{eqnarray}
where $c=\frac{b}{a}$, and 
\begin{eqnarray}
\left\langle (-1)^{j}\tau _{j}^{x}\right\rangle &=&\frac{1}{2}\cos \left(
\gamma \right) \\
\left\langle \tau _{j}^{z}\right\rangle &=&\frac{1}{2}\sin \left( \gamma
\right)  \nonumber
\end{eqnarray}
The general character of the phase transition in the Hamiltonian (\ref{H-CI}%
) is well known \cite{soliton-lattice}: The system remains commensurate
until $|\delta |$ exceeds a finite critical value $\delta ^{c}$. Therefore,
the {\em Toulouse limit is proved to be stable over a finite range of
parameter space}{\it , }$\Delta J_{z}\neq 0${\it .}

Yet, care should be taken to identify the exact nature of the transition and
the character of the ensuing gapless phase. The groundstate of (\ref{H-CI})
is determined by the field configuration that minimize the energy. As we
shall see, the C-I transition in our Hamiltonian (\ref{H-MF}) is unusual due
to the fact that the parameters $\delta $ and $h$ are themselves not
constants, but instead are dynamic fields which need to be {\em determined
self-consistently by the additional mean-field minimization condition} on $%
\left\langle \tau _{j}^{z}\right\rangle $ and $\left\langle \tau
_{j}^{x}\right\rangle $. There are, in principle, three possible ground
state solutions for the Hamiltonian (\ref{H-CI}):

\begin{itemize}
\item  Phase-1: A uniform spin gap phase, identical to the Toulouse point
solution, with no finite gradients of $\partial _{x}\phi _{s}(x_{j})$ i.e., $%
\left\langle \cos [\sqrt{2\pi }\phi _{s}(j)]\right\rangle \neq 0$, $%
\left\langle (-1)^{j}\tau _{j}^{x}\right\rangle \neq 0$, $\left\langle
\partial _{x}\phi _{s}\right\rangle =0$, and hence, also $\left\langle \tau
_{j}^{z}\right\rangle =0$.

\item  phase-2: A gapless incommensurate spin ''soliton lattice'' ground
state with periodic step-like kinks in the $\phi _{s}(x_{j})$ field. In such
a phase, $\left\langle \partial _{x}\phi _{s}\right\rangle \neq 0$, but
still $\left\langle \cos [\sqrt{2\pi }\phi _{s}(j)]\right\rangle \neq 0$,
and both $\left\langle (-1)^{j}\tau _{j}^{x}\right\rangle \neq 0$ and $%
\left\langle \tau _{j}^{z}\right\rangle \neq 0$.

\item  Phase-3: A free gapless SDW phase; $\left\langle \partial _{x}\phi
_{s}\right\rangle \neq 0$, $\left\langle \cos [\sqrt{2\pi }\phi
_{s}(j)]\right\rangle =0$. In that phase $\left\langle \tau
_{j}^{x}\right\rangle =0$ and $\left\langle \tau _{j}^{z}\right\rangle \neq
0 $.
\end{itemize}

The name ''soliton lattice'' comes from the classical solution, which has
long range periodic order. Quantum fluctuations turn the long range order
into power-law correlations, and thus the quantum ground state should
properly be termed a soliton liquid. Nevertheless, this does not change the
qualitative distinctions (in terms of non-zero expectation values) between
the various phases. For simplicity, I will {\em discuss the phases in
classical terms}.

To find the transition points between the phases we need to compare, for a
given $\Delta J_{z}=\left( J_{z}-\pi v_{F}\right) \neq 0$, the ground state
energies of the spin gap phase (phase-1) with the that of the gapless
phases. The usual result for commensurate-incommensurate transition, where
the parameters $h$ and $\delta $ are constant, is that the soliton lattice
solution (phase-2) has lower energy than the SDW solution, and transition is
second order. This is {\em not} the case here, due to the fact that the {\em %
parameters }$h${\em \ and }$\delta ${\em \ are themselves inter-dependent
dynamic variables. Thus, we need to minimized the ground state energy with
respect to both the soliton spacing, }$l${\em , (as usually done) and also
with respect to the mean-field parameter}, $\gamma $.

The resulting commensurate-incommensurate transition in the transformed 1D
Kondo lattice Hamiltonian (\ref{H-CI}), is first order. The argument is the
Following: Remember that $\tau _{j}^{z}$ and $\tau _{j}^{x}$ are
non-commuting. Therefore, if there was a second order transition to the
soliton lattice phase than at the transition point, $\delta \approx \frac{4}{%
\pi }\sqrt{h_{0}}$, both $\delta \sim \left\langle \tau
_{j}^{z}\right\rangle \neq 0$ and $h\sim \left| \left\langle \tau
_{j}^{x}\right\rangle \right| \neq 0$ are less by a {\em finite} amount from
their respective maximum value, $\delta _{0}$ and $h_{0}$. The energy of the
soliton lattice at the second order transition is equal to the energy of the
commensurate phase with the same value of $h$, which is always less than the
maximum energy of the commensurate phase-1 (for which $h=h_{0}$ and $\delta
=0$). Thus, we establish that the commensurate-incommensurate transition is
necessarily first order.

But, what is the incommensurate phase? There is no closed expression for the
soliton lattice energy away from the dilute limit (i.e., far from the
putative second order transition). Yet, we can analyze the competition
between phase-2 and phase-3 in the dense soliton lattice limit (when the
distance between soliton centers is less than the single soliton width). In
that limit, the commensurate energy contribution (due to $h\neq 0$) is
exponentially small, while the $\partial _{x}\phi _{s}$ term contribution is
approximately linear in $\delta <\delta _{0}$. Thus (again in contrast with
the usual case of constant coefficients $\delta $,$h\neq 0$), the dense
soliton lattice energy is less favorable than the $SDW$ phase-3, (in which $%
h=0$ and $\delta =\delta _{0}$).

The above argument leads to two possible scenarios: Either there is a
sequence of two first order transitions (phase-1$\longrightarrow $phase-2$%
\longrightarrow $phase-3). Or, there is one first order transition (phase-1$%
\longrightarrow $phase-3). We conjecture that the second possibility is the
correct one, and hence the phase transition occurs at $\delta
_{0}^{critical}=\sqrt{2h_{0}}$, i.e., 
\begin{equation}
\left( \Delta J_{z}\right) _{critical}=\sqrt{\frac{J_{\perp }}{2v_{s}a}}
\end{equation}
.

In conclusion, at a finite deviation $\left( \Delta J_{z}\right) _{critical}$
from the Toulouse point towards strong coupling, there is a{\em \
first-order commensurate-incommensurate transition in the spin field }$\phi
_{s}$, {\em in conjunction with transformed impurity spin flop transition} 
{\em from} $\left\{ \left\langle \tau _{j}^{x}\right\rangle \neq
0,\left\langle \tau _{j}^{z}\right\rangle =0\right\} $\ {\em to} $\left\{
\left\langle \tau _{j}^{x}\right\rangle =0,\left\langle \tau
_{j}^{z}\right\rangle \neq 0\right\} $. The transition\ is from the spin gap
phase-1, to the gapless $SDW$\ phase-3, with no soliton lattice region.

\subsection{Staggered Luttinger liquid - a new strong coupling phase}

The transition in the spin sector to the gapless $SDW$ phase leads to a new
state which we call ''staggered-Luttinger-liquid''. The staggered-LL is
expressed in terms of the transformed fermion fields, which have composite
phase fields, 
\begin{eqnarray}
\tilde{R}_{\sigma }(x) &\equiv &UR_{\sigma }(x)U{^{\dag }}=R_{\sigma
}(x)e^{+i2\pi \sum_{x_{j}<x}\tau _{j}^{z}\sigma } \\
\tilde{L}_{\sigma }(x) &\equiv &UL_{\sigma }(x)U{^{\dag }}=L_{\sigma
}(x)e^{-i2\pi \sum_{x_{j}<x}\tau _{j}^{z}\sigma }
\end{eqnarray}
where $\sigma =\pm \frac{1}{2}$ is the electron spin, and $\tau ^{z}$ is the
impurity operator which can take values $\pm \frac{1}{2}$, (so $2\pi \tau
^{z}\sigma =\pm \frac{\pi }{2}$).

Same as for the Toulouse point, we calculate the correlation functions with
respect to the spectrum of the transformed Hamiltonian. All the order
parameters which in Bosonic form depend on the $\phi _{s}$ field, have {\em %
staggered} correlation functions (as defined in (\ref{stagger})) {\em %
irrespective of the impurity configuration} $\left\{ \tau _{j}^{z}\right\} $%
. Since there is no spin gap, there are now also gapless spin density wave
(SDW) and triplet pairing modes. In the bosonization representation, both $%
\cos \left( \sqrt{2\pi }\phi _{s}\right) $ and $\sin \left( \sqrt{2\pi }\phi
_{s}\right) $ have power law decay of correlations (with an added staggered
factor $\left( -1\right) ^{j-j^{\prime }}$)

The $\tau _{j}^{z}$ order of the transformed impurity array requires further
clarification. The inter-impurity interactions generated by integrating out
the transformed 1DEG degrees of freedom in the residual Kondo interaction, $%
\Delta J_{z}^{f}\sum_{j}\tau _{j}^{z}\partial _{x_{j}}\phi _{s}$, are long
ranged (i.e., well beyond nearest neighbor interaction). Honner and Gulacsi 
\cite{Gulasci-(Kondo-ferro)} suggest that the effective interaction is
ferromagnetic, and thus at least conforms with strong coupling calculations 
\cite{Ueda-PhaseDiagram}.

\subsection{staggered-BCS; A third spin gap phase?}

It is interesting to investigate what would be the form of a BCS pairing of
a composite staggered-LL? i.e., we introduce a conventional weak attractive
interaction, $U<0$ (e.g., due to phonons), to the 1DEG Hamiltonian. The
singlet pairing take the form 
\begin{eqnarray}
\tilde{O}_{SP} &=&\frac{1}{\sqrt{2}}\left[ \tilde{L}_{\uparrow }\tilde{R}%
_{\downarrow }+\tilde{R}_{\uparrow }\tilde{L}_{\downarrow }\right]
\label{staggered-BCS} \\
&=&\left( -1\right) ^{j\left( x\right) }\frac{1}{\sqrt{2}}\left[ L_{\uparrow
}R_{\downarrow }+R_{\uparrow }L_{\downarrow }\right] =\left( -1\right)
^{j\left( x\right) }{\cal O}_{SP}.  \nonumber
\end{eqnarray}
The resulting pair correlation function is staggered (as defined in (\ref
{stagger})), with nodes at the Kondo impurity periodicity. It corresponds to
a negative Josephson coupling across each Kondo impurity\cite
{Zachar-KondoStrong}. We stress that the node in the pair correlation
function due to negative Josephson coupling is a {\em node in the pair
center-of-mass motion}. It should not be confused with a node in the
relative pair state\cite{Varma1987(Kondo)}. There is {\em no} gapless $k=0$
pairing mode.

On the other hand, {\em all} the composite modes are now incoherent! In
order to see this, note that in the gapless staggered LL phase, the gapless
composite pairing mode ${\cal O}_{c-SP}$ came from the component $\sin
\left( \sqrt{2\pi }\phi _{s}\right) \tau _{j}^{z}$ (see eq.\ref{APX O_c-SP}
in the appendix). Due to the singlet pairing interaction $\left\langle \cos
\left( \sqrt{2\pi }\phi _{s}\right) \right\rangle \neq 0$, and thus the
correlation function $\left\langle \sin \left( \sqrt{2\pi }\phi _{s}\left(
x\right) \right) \sin \left( \sqrt{2\pi }\phi _{s}\left( x^{\prime }\right)
\right) \right\rangle $ is exponentially decaying. Moreover, as in the
staggered LL, $\left\langle \tau _{j}^{z}\right\rangle \neq 0$ and thus the
part $\left( R_{\uparrow }^{\dagger }L_{\uparrow }^{\dagger }\tau
^{-}-R_{\downarrow }^{\dagger }L_{\downarrow }^{\dagger }\tau ^{+}\right) $
of ${\cal O}_{c-SP}$ is also exponentially decaying. These results are
summarized in line 3 of table-1.

Our analysis suggests a new possibility: An {\em unconventional}
staggered-BCS pairing phase may arise out of a two step process, where the
staggered-LL is a precursor to the staggered-BCS phase; First, at a
temperature $T_{hf}$, set by the renormalized Kondo interaction, there is a
cross-over to a staggered-LL phase, characterized by the unitarity limit
phase shifts. Then, at a much lower temperature, $T_{c}$, a {\em conventional%
} BCS pairing mechanism (e.g., phonons) leads to the un-conventional finite
momentum BCS pairing state. The above demonstrates the importance of
considering the cross-over effects, due to strong interactions, prior to the
consideration of pairing mechanisms.

\section{Concluding remarks}

The main results of this paper are: (1) The identification of the staggered
liquids family of fixed point, as summarized in tables-1,2. (2) Derivation
of the phase transition from a spin gap phase at intermediate coupling to a
gapless staggered LL at strong coupling. (3) The commutation relations (\ref
{[CDW,eta-even]}-\ref{[c-CDW,eta-odd]}) which relate CDW and pairing modes.
Below we make some additional comments on our results.

At weak coupling, the Kondo-Heisenberg model consists of a free electron gas
coupled to a spin density wave system. One would naturally expect a BCS
mechanism leading to a state of $k=0$ BCS pairing of conduction electrons
mediated by spin waves of the Heisenberg chain. We find it quite surprising
that such a state does not materialize at any stable fixed point of the one
dimensional problem.

Previous numerical simulations in the strong coupling limit\cite
{Ueda-PhaseDiagram} have found that the ''dominant'' gapless CDW mode (in a
gapless strong coupling LL phase) is with ''large Fermi sea'' wavenumber $%
2k_{F}^{\ast }$. Pairing modes were never evaluated. Yet, from our
commutation relations (\ref{[CDW,eta-even]},\ref{[c-CDW,eta-odd]}) it is
imperative that the pairing correlations are staggered!! We comment that,
following the analysis in this paper, it is important that numerical
simulations will establish the {\em existence} of both ${\cal O}_{CDW}$ and $%
{\cal O}_{c-CDW}$ gapless CDW modes. The pairing modes then follow
automatically as we explained.

The numerical simulations were performed in the extreme strong coupling
limit on a particular lattice structure for which our analytical methods are
not rigorously valid. Therefore, it is important to establish whether the
gapless strong coupling LL phase in the numerical simulations is identical
to the one which we derived analytically by a phase transition from the
Toulouse limit solution\cite{ZacharTsvelik-KondoHeisenberg}.

\begin{acknowledgement}
: I thank Steve Kivelson and Thierry Giamarchi for stimulating discussions.
I thank the hospitality of the ITP at Santa Barbara NSF grant PHY999-07949,
and Jacquelin Mansard on the north shore of Oahu where part of this work was
completed.
\end{acknowledgement}

\section{Appendix: Discussion of order parameters}

The study of the different stable phases of the Kondo-Heisenberg array
begins with an analysis of the gapless excitations of the decoupled fixed
point. From there, as usual, we sort the phases by determining which of
these excitations become gapped, and which remain gapless in the presence of
the (Kondo) couplings between the 1DEG and the Heisenberg chain. In order to
facilitate the readability of the paper, we give below the explicit
expressions of various order parameters:

\subsubsection{Density wave modes}

The low energy spin currents of the 1DEG, $\vec{s}\left( x\right) $, can be
decomposed into two parts; 
\begin{equation}
\vec{s}\left( x\right) =\vec{J}_{s}\left( x\right) +\left[ \vec{n}_{s}\left(
x\right) e^{i2k_{F}x}+{\rm H.c.}\right]
\end{equation}
where 
\begin{eqnarray}
\vec{J}_{s} &=&\sum_{\lambda ,\sigma ,\sigma ^{\prime }}\psi _{\lambda
,\sigma }^{\dagger }\frac{\vec{\sigma}_{\sigma ,\sigma ^{{\prime }}}}{2}\psi
_{\lambda ,\sigma ^{\prime }} \\
\vec{n}_{s} &=&\sum_{\sigma ,\sigma ^{\prime }}R_{\sigma }^{\dagger }\frac{%
\vec{\sigma}_{\sigma ,\sigma ^{{\prime }}}}{2}L_{\sigma ^{\prime }}
\end{eqnarray}
are respectively the $k=0$ and the $k=2k_{F}$ components of the SDW mode
(charge-$0$, spin-$1$) of the 1DEG. (The index $\lambda =R,L$ correspond to
Right/Left going electron fields).

The Heisenberg chain spin current, $\vec{\tau}_{j}$, may be similarly
decomposed into a $k=0$ part, $\vec{J}_{\tau }$, and a finite momentum $k=%
\frac{\pi }{b} $ part, $\left( -1\right) ^{j}\vec{n}_{\tau }$ (where $\frac{%
2\pi }{b}$ is the reciprocal lattice vector of the Heisenberg chain); 
\begin{equation}
\vec{\tau}_{j}=\vec{J}_{\tau }\left( x_{j}\right) +\left( -1\right) ^{j}\vec{%
n}_{\tau }\left( x_{j}\right)
\end{equation}

For the density-wave excitations, we count only the number of finite
momentum excitations. It follows by symmetry that, for finite momentum, if
there is a gapless mode at momentum $q$ than there is also a gapless mode
with momentum $-q$. We count them as one mode. To summarize, the gapless
spin-1 excitations of the 1DEG and the Heisenberg spin chain, and the
operator whose correlation function is most directly sensitive to it are
listed in table-A1. 
\begin{equation}
\stackrel{Table-A1:\text{ Gapless SDW excitations}}{
\begin{tabular}{||c|c||}
\hline\hline
$
\begin{array}{c}
operator
\end{array}
$ & $
\begin{array}{c}
wave \\ 
number
\end{array}
$ \\ \hline
$\vec{n}_{s}$ & $2k_{F}$ \\ \hline
$\vec{n}_{\tau }$ & $\frac{\pi }{b}$ \\ \hline\hline
\end{tabular}
}
\end{equation}

The incommensurate 1DEG has one CDW excitation (charge-$0$ spin-$0$) with
momentum, $2k_{F}$, created by the operator 
\begin{eqnarray}
{\cal O}_{CDW} &=&\frac{1}{2}\sum_{\lambda ,\sigma }\psi _{\lambda ,\sigma
}^{\dagger }\psi _{-\lambda ,\sigma }.  \label{APX    O_CDW} \\
&\sim &e^{+i\left[ \sqrt{2\pi }\phi _{c}+2k_{F}x\right] }\cos \left( \sqrt{%
2\pi }\phi _{s}\right)
\end{eqnarray}
The generalized Luttinger theorem\cite{Generalized-Luttinger} asserts that
there must be a gapless CDW mode at $2k_{F}^{\ast }=2k_{F}+\frac{\pi }{b}$.
It is realized by the existence of composite-CDW\cite{Col-Tsv-Geo} order
parameters which are formed by combining a spin-$1$ SDW of the 1DEG with a
spin-$1$ SDW of the Heisenberg chain into a composite singlet $\hat{O}%
_{c-CDW}$, 
\begin{eqnarray}
{\cal O}_{c-CDW} &=&\vec{s}\cdot \vec{\tau}  \nonumber \\
&=&\vec{n}_{R}\cdot {\bf \tau =}\frac{1}{2}\left( n^{+}\tau
_{j}^{-}+n^{-}\tau _{j}^{+}\right) +n^{z}\tau _{j}^{z} \\
&=&\vec{J}_{s}\cdot \vec{J}_{\tau }+\vec{J}_{s}\cdot \vec{n}_{\tau }\left(
-1\right) ^{j}  \nonumber \\
&+&\left[ e^{i2k_{F}x}\vec{n}_{s}\cdot \vec{J}_{\tau }+h.c.\right]  \nonumber
\\
&+&\left[ e^{i2k_{F}x}\vec{n}_{s}\cdot \vec{n}_{\tau }+h.c.\right] \left(
-1\right) ^{j}.  \label{O_c-CDW}
\end{eqnarray}
To summarize, the non-interacting two-chain system of a Luttinger liquid and
a Heisenberg spin chain has {\em gapless finite momentum CDW modes at three
wave vectors} (table-A2). 
\begin{equation}
\stackrel{Table-A2:\text{ Gapless CDW excitations}}{
\begin{tabular}{||l|l||}
\hline\hline
$
\begin{array}{c}
operator
\end{array}
$ & $
\begin{array}{c}
wave \\ 
number
\end{array}
$ \\ \hline
$\vec{n}_{s}\cdot \vec{n}_{\tau }$ & $2k_{F}+\frac{\pi }{b}$ \\ \hline
${\cal O}_{CDW}$ & $2k_{F}$ \\ \hline
$\vec{n}_{s}\cdot \vec{J}_{\tau }$ & $2k_{F}$ \\ \hline
$\vec{J}_{s}\cdot \vec{n}_{\tau }$ & $\frac{\pi }{b}$ \\ \hline\hline
\end{tabular}
}
\end{equation}
Note that the composite-CDW excitations at wave vectors $\frac{\pi }{b}$ and 
$2k_{F}+\frac{\pi }{b}$ are not independent, since they can be related
through a multiplication by the 1DEG ${\cal O}_{CDW}$ (which has wave vector 
$2k_{F}$). Thus, there are only three independent gapless CDW modes.

\subsubsection{Singlet pairing modes}

The charge-$2e$ singlet pairing modes also require careful consideration. In
addition to the usual $k=0$ BCS even parity singlet pairing, 
\begin{eqnarray}
{\cal O}_{SP} &=&\frac{1}{\sqrt{2}}\left[ L_{\uparrow }^{\dagger
}R_{\downarrow }^{\dagger }+R_{\uparrow }^{\dagger }L_{\downarrow }^{\dagger
}\right]  \label{APX   O_sp} \\
&\sim &e^{+i\sqrt{2\pi }\theta _{c}}\cos \left( \sqrt{2\pi }\phi _{s}\right)
\nonumber
\end{eqnarray}
we note also the existence of an $\eta $-pairing mode at momentum $\pm
2k_{F} $, 
\begin{eqnarray}
\eta _{R} &=&R_{\uparrow }^{\dagger }R_{\downarrow }^{\dagger }\sim e^{+i%
\sqrt{2\pi }\theta _{c}}e^{-i\left[ \sqrt{2\pi }\phi _{c}+2k_{F}x\right] } \\
\eta _{L} &=&L_{\uparrow }^{\dagger }L_{\downarrow }^{\dagger }\sim e^{+i%
\sqrt{2\pi }\theta _{c}}e^{+i\left[ \sqrt{2\pi }\phi _{c}+2k_{F}x\right] } 
\nonumber
\end{eqnarray}
corresponding to right and left going singlet pairs.

As with the CDW modes, in addition to the singlet pairing modes of the 1DEG,
it is necessary to consider the {\em composite} singlet pairing, $O_{c-SP}$,
(a product of a triplet pairing in the 1DEG with a spin-$1$ mode of the
Heisenberg chain) which turns out to be {\em odd parity}\cite
{odd-w,Coleman-Miranda}, 
\begin{eqnarray}
{\cal O}_{c-SP}= &&-i\frac{1}{2}\left( R^{\dagger }{\bf \vec{\sigma}}\sigma
_{2}L^{\dagger }\right) \cdot {\bf \vec{\tau}}  \label{APX O_c-SP} \\
= &&\frac{1}{2}\left[ \left( R_{\uparrow }^{\dagger }L_{\uparrow }^{\dagger
}\tau _{j}^{-}-R_{\downarrow }^{\dagger }L_{\downarrow }^{\dagger }\tau
_{j}^{+}\right) \right.   \nonumber \\
&&\;\;\;\;\;\;\;\;\;\;\;\;\;\;\;\;\;\;\;\;\;\;\;\;\;\;\;\;\;\;\;\;\;\;\;\;%
\left. -\left( R_{\uparrow }^{\dagger }L_{\downarrow }^{\dagger
}+R_{\downarrow }^{\dagger }L_{\uparrow }^{\dagger }\right) \tau _{j}^{z}
\right]   \nonumber \\
\sim  &&e^{+i\sqrt{2\pi }\theta _{c}}\left[ e^{-i\sqrt{2\pi }\theta
_{s}}\tau _{j(x)}^{+}+e^{+i\sqrt{2\pi }\theta _{s}}\tau _{j(x)}^{-}\right.  
\nonumber \\
&&\;\;\;\;\;\;\;\;\;\;\;\;\;\;\;\;\;\;\;\;\;\;\;\;\;\;\;\;\;\;\;\;\;\;\;\;%
\left. +2i\sin \left( \sqrt{2\pi }\phi _{s}\right) \tau ^{z}\right]  
\nonumber
\end{eqnarray}
(Note: If we do not take the Klein factors carefully into account than the
bosonized form of the singlet and triplet composite pairing is erroneously
exchanged!). It can be decomposed into two momentum components: a uniform $%
k=0$ composite singlet 
\begin{equation}
\hat{O}_{c-SP}^{k=0}\left( x\right) =-i\frac{1}{2}\left( R^{\dagger }{\bf 
\vec{\sigma}}\sigma _{2}L^{\dagger }\right) \cdot \vec{J}_{\tau }
\label{k=0 c-sp}
\end{equation}
and a $k=\frac{\pi }{b}$ , {\it i.e.} a {\em staggered} composite singlet 
\begin{equation}
\hat{O}_{c-SP}^{stagger}\left( x\right) =-i\frac{1}{2}\left( R^{\dagger }%
{\bf \vec{\sigma}}\sigma _{2}L^{\dagger }\right) \cdot \vec{n}_{\tau }\left(
-1\right) ^{j}  \label{staggered- c-sp}
\end{equation}
The commutation relations (\ref{[CDW,eta-even]}-\ref{[c-CDW,eta-odd]})
relate to each gapless CDW mode a corresponding gapless pairing mode.
Therefore, formally, only the $\eta $-pairing modes need to be counted. The
concomitant ``trivial'' existence of the usual BCS pairing ${\cal O}_{SP}$
and composite pairing ${\cal O}_{c-SP}$ mode should be implicitly understood.

The operator $O_{c-SP}$ is odd under spin inversion operation ($R_{\uparrow
}^{\dagger }\leftrightarrow R_{\downarrow }^{\dagger }$, $\tau
^{-}\leftrightarrow \tau ^{+}$, $\tau ^{z}\leftrightarrow -\tau ^{z}$), as
expected for a singlet. Note that it's spin inversion parity is odd, even
though the conduction electrons part is in triplet pairing. In that sense
the order parameter is a composite singlet. The operator is clearly odd
under space inversion operation, $P$ (exchanging $R$ and $L$). The composite
singlet operator, $O_{c-SP}$, can be arrived at by taking the time
derivative of the BCS singlet order parameter \cite{odd-w}, ${\frac{\partial
O_{SP}}{\partial t}}\propto \left[ H_{K},O_{SP}\right] =O_{c-SP}$, where $%
H_{K}$ is the Kondo-Heisenberg Hamiltonian (\ref{H-Kondo}). Therefore, $%
O_{c-SP}$ is {\sl odd} under time reversal, or alternatively, has only $%
odd-w $ dependence. The corresponding order parameter on a discrete lattice
(e.g. on a zigzag ladder) is \cite{coleman(my-paper)} 
\[
O_{c-SP}=-\frac{i}{2}\left( -1\right) ^{j}\left( \psi _{j}^{\dagger }{\bf %
\sigma }\sigma _{2}\psi _{j+1}^{\dagger }\right) \cdot {\bf \tau }_{j} 
\]
(The factor $\left( -1\right) ^{j}$ is needed so that both odd and even $%
j-sites$ will conform in the continuum limit representation).

There is a qualitative difference between the commutation relation (\ref
{[CDW,eta-odd]}) and previous commutation relations of ${\cal O}_{c-SP}$ in
the literature. As elaborated below, we generated the composite singlet, $%
O_{c-SP}$, by a combination of $2k_{F}^{\ast }$ composite particle-hole mode
($\vec{n}_{R}\cdot {\bf \tau }$) and {\sl finite momentum} $k=2k_{F}$
singlet ($\eta _{L}$ pairing): 
\begin{eqnarray}
\vec{n}_{R} &=&\vec{O}_{2k_{F}-SDW}=R_{\alpha }^{+}\frac{\vec{\sigma}%
_{\alpha \beta }}{2}L_{\beta }  \nonumber \\
O_{c-CDW} &=&\vec{n}_{R}\cdot {\bf \tau =}\frac{1}{2}\left( n^{+}\tau
_{j}^{-}+n^{-}\tau _{j}^{+}\right) +n^{z}\tau _{j}^{z}  \nonumber \\
O_{c-SP} &=&\left[ \eta _{L},O_{c-CDW}\right]
\end{eqnarray}

Note that the above generation of the composite singlet pairing is different
from the usual way in which it is generated \cite{odd-w} using the $\frac{%
\pi }{b}$ momentum composite particle-hole mode ($\vec{J}_{R}\cdot {\bf \tau 
}$) and $k=0${\sl \ }momentum singlet ($O_{SP}$ pairing): 
\begin{eqnarray}
\vec{J}_{R} &=&R_{\alpha }^{+}\frac{\vec{\sigma}_{\alpha \beta }}{2}R_{\beta
}  \nonumber \\
{\cal O}_{SP} &=&\frac{1}{\sqrt{2}}\left[ R_{\uparrow }^{\dagger
}L_{\downarrow }^{\dagger }-R_{\downarrow }^{\dagger }L_{\uparrow }^{\dagger
}\right]  \nonumber \\
{\cal O}_{c-SP} &=&\left[ {\cal O}_{SP},\vec{J}_{R}\cdot {\bf \tau }\right]
\end{eqnarray}
$\vec{J}_{R}\cdot {\bf \tau }$ is an interaction term in the Hamiltonian,
which develops a non-zero expectation value, $\left\langle \vec{J}_{R}\cdot 
{\bf \tau }\right\rangle \neq 0$, in the spin gap phase of the Kondo lattice
Hamiltonian. Thus, the relation $O_{c-SP}=\left[ O_{SP},\vec{J}_{R}\cdot 
{\bf \tau }\right] =\left[ O_{SP},H\right] $ is important for establishing
the time reversal symmetry of $O_{c-SP}$ as determined by the Hamiltonian.
The $\left( \vec{J}_{R}\cdot {\bf \tau }\right) $ operator is {\em not one
of the gapless modes}. In contrast, $O_{c-CDW}=\vec{n}_{R}\cdot {\bf \tau }$
is a gapless mode in the spin gap phase. Thus, our relation $O_{c-SP}=\left[
\eta _{L},O_{c-CDW}\right] $, is establishing the {\em inter-dependence of
gapless modes} in the spin gap phase.

The commutation relations (\ref{[c-CDW , eta-even]}) and (\ref{[CDW ,
eta-even]}) indicate that there must be some symmetry difference between the
usual CDW ($O_{CDW}$) mode of the 1DEG and the composite-CDW ($O_{c-CDW}$)
mode, and that the composite-CDW cannot be used in combination with $\eta
^{even}$to construct a BCS singlet mode, $\Delta $, (as can be done with the
usual CDW). Clearly there is no difference in the global symmetry properties
of the two CDW modes (this would have been a violation of the generalized
Luttinger's theorem). The difference is in a {\em relative internal symmetry}
of the two chain system; a $\pi $ relative spin rotation around the $z-axis$
of the 1DEG with respect to the Heisenberg spin chain. This effect is best
seen from the Bosonized spin field dependence of the operators, 
\begin{eqnarray*}
\Delta &\sim &\cos \left( \sqrt{2\pi }\phi _{1s}\right) \\
O_{CDW} &\sim &\cos \left( \sqrt{2\pi }\phi _{1s}\right) \\
O_{c-CDW} &\sim &\cos \left( \sqrt{2\pi }\left[ \theta _{1s}-\theta _{2s}%
\right] \right)
\end{eqnarray*}
(where subscripts ''1'' and ''2'' refer to the 1DEG and the impurity spin
chain respectively). A $\pi $ relative spin rotation around the $z-axis$ is
shifting $\sqrt{2\pi }\left[ \theta _{1s}-\theta _{2s}\right] $ by $\pi $
and leaving $\phi _{1s}$ unaffected.. Thus under this operation, which we
label ${\cal R}_{z}^{S-rel}\left( \pi \right) $, 
\begin{eqnarray}
\eta &\longrightarrow &+\eta  \label{Zrot-symmetry} \\
\Delta &\longrightarrow &+\Delta  \nonumber \\
O_{CDW} &\longrightarrow &+O_{CDW}  \nonumber \\
O_{c-CDW} &\longrightarrow &-O_{c-CDW}.  \nonumber
\end{eqnarray}
From these transformation properties (\ref{Zrot-symmetry}) it is clear that
the composite-CDW cannot be used in combination with $\eta ^{even}$to
construct a BCS singlet mode, since under ${\cal R}_{z}^{S-rel}\left( \pi
\right) $ , $\Delta \longrightarrow +\Delta $, while $\eta
^{even}O_{c-CDW}\longrightarrow -\eta ^{even}O_{c-CDW}$. Hence, our final
conclusion is that the Toulouse point phase and the weak coupling limit spin
gap phase of the Kondo-Heisenberg model are distinct phases (as summarized
in table-1).

There is a simple physical interpretation for what is the distinction made
by the ${\cal R}_{z}^{S-rel}\left( \pi \right) $ symmetry operation: The
composite-CDW ($O_{c-CDW}$) is actually constructed out of two spin-1 SDW
modes which are coherently combined into a total spin singlet. Therefore,
the mode is sensitive to the coherent relative phases of the spin fields
between the 1DEG and the Heisenberg chain, which is probed by the ${\cal R}%
_{z}^{S-rel}\left( \pi \right) $. In contrast, the ''pure'' CDW mode ($%
O_{CDW}$) of the 1DEG is independent of any relative state of the Heisenberg
chain.

\end{document}